\documentclass[aps,twocolumn,showpacs,showkeys]{revtex4}
\usepackage{graphicx}

\begin{document}


\newcommand\beq{\begin{equation}}
\newcommand\eeq{\end{equation}}
\newcommand\beqa{\begin{eqnarray}}
\newcommand\eeqa{\end{eqnarray}}
\newcommand\ket[1]{|#1\rangle}
\newcommand\bra[1]{\langle#1|}
\newcommand\scalar[2]{\langle#1|#2\rangle}

\newcommand\jo[3]{\textbf{#1}, #3 (#2)}


\title{\Large\textbf{Unconditionally secure bit commitment by causally independent encryptions}}

\author{Chi-Yee Cheung}

\email{cheung@phys.sinica.edu.tw}

\affiliation{Institute of Physics, Academia Sinica\\
             Taipei, Taiwan 11529, Republic of China\\}


\begin{abstract}
We propose a new classical bit commitment protocol using the relativistic constraint that signals cannot travel faster than the speed of light $c$. This protocol is unconditionally secure against both classical or quantum attacks. The sender (Alice) and the receiver (Bob) each controls two secure stations separated by a large distance $d$, and they communicate by exchanging classical information only. Alice commits by sending from her stations two causally independent encrypted messages to the neighboring Bob's stations, after that the protocol is out of her control and she plays no role in the unveiling phase. The commitment remains concealed for a period of $\Delta t=d/2c$. This protocol requires only limited communication resources and is readily implementable with current technologies.
\end{abstract}



\maketitle


Cryptography is a science with important practical applications. The most well known task in cryptography is key distribution whose security can be guaranteed by quantum mechanical means \cite{BB84,Stucki2009}. Bit commitment is another basic cryptographic task that has attracted much interest. The security of bit commitment is an important issue because, apart from being useful in its own right, it can be used as a building block for other more complicated cryptographic tasks such as coin tossing \cite{Blum83}, oblivious transfer \cite{Bennett91,Crepeau94,Yao95,Mayers96}, zero-knowledge proofs \cite{Brassard88}, and secure two-party computation \cite{Kilian88,Crepeau95}, etc.

A bit commitment protocol involves two distrusting parties (Alice and Bob), and it consists of two separate procedures, {\it Commit} and {\it Unveil}. In {\it Commit}, Alice commits to Bob a secret bit $b\in\{0,1\}$ and provides him with a piece of evidence to bind her commitment. In {\it Unveil}, Alice reveals the value of $b$ and normally needs to provide additional information to Bob so that he can use the evidence to check if the unveiled bit value is indeed the same as what she had committed before. A bit commitment protocol is secure if it is both concealing and binding. Concealing means Bob does not know the value of $b$ before Alice reveals it, and binding means Alice cannot change the committed value of $b$ without the risk of being discovered by Bob. Furthermore if the protocol remains secure even if Alice and Bob had capabilities limited only by the laws of nature (this is sometimes referred to as having unlimited computational power), then it is said to be unconditionally secure.

A typical example of classical commitment is for Alice to write the committed bit on a piece of paper and lock it in a safe. Alice gives the locked safe to Bob as the evidence but keeps the key. Later Alice unveils the bit and provides the key with which Bob can check if Alice is honest.
However, like classical cryptography, bit commitment based on the exchange of classical information alone cannot be unconditionally secure. The securities of all such protocols are vulnerable to advances in technologies and/or breakthroughs in solving computationally hard tasks.  With the success of quantum key distribution in cryptography \cite{BB84}, there was initially high hope that one could also use the laws of quantum physics in bit commitment to guarantee its security \cite{BCJL93,Brassard96}. Unfortunately this expectation turned out to be unfounded. In 1997, Mayers \cite{Mayers97} and Lo and Chau \cite{LoChau97} put forth a ``no-go theorem" showing that all concealing quantum bit commitment protocols are not binding. The proof is based on the observation that, using quantum entanglement, the sender Alice can keep all undisclosed classical information undetermined and stored at the quantum level. In particular, she can choose to delay all prescribed measurements without consequences until it is required to disclose the outcomes. Then it can be shown mathematically that Alice can unveil $b=0$ or $b=1$ as she wishes. Despite the no-go theorem, the subject of unconditionally secure quantum bit commitment has continued to attract considerable interests. Some reinforced the impossibility arguments \cite{Brassard97,Bub01,Cheung07,DAriano07,Chiribella13,Magnin10}, while others tried to show the converse \cite{Yuen08,Cheung09,He06,He11}. However, without introducing new ingredients into the problem, so far no one has been able to construct a quantum bit commitment protocol which can be convincingly shown to be unconditionally secure.

The no-go theorem of Mayers-Lo-Chau is based implicitly on non-relativistic physics which is only an approximation in a world where the speed of light can be taken to be infinity. The principle of relativity however dictates that nothing can travel faster than the speed of light. Using this constraint, Kent has successfully constructed a classical bit commitment scheme which can be shown to be unconditionally secure \cite{Kent99}. Apart from using relativistic physics, this protocol departs from the standard bit commitment scenario in that Alice ($A$) and Bob ($B$) are allowed to set up trusted stations located far away from themselves. This additional feature is also present in all other versions of relativistic bit commitment protocols proposed in recent years \cite{Kent11,Kent12,Croke12}.

In the latest 2012 proposal \cite{Kent12}, Alice and Bob each control three secure stations, ($A_1$,$A_2$,$A_3$) and ($B_1$,$B_2$,$B_3$), respectively. The sender stations $A_2$ and $A_3$ are located far away on opposite sides of $A_1$, and each $A_i$ has a receiver station $B_i$ located adjacent to it. $B_1$ sends $A_1$ a number of randomly chosen BB84 qubits \cite{BB84}, depending on the value of $b$ she wants to commit, $A_1$ measures the qubits in one of the two BB84 bases and sends the measurement outcomes securely to $A_2$ and $A_3$, who then unveil by revealing the outcomes to $B_2$ and $B_3$ respectively. This protocol is experimentally feasible and has been successfully implemented recently \cite{Lunghi13,Liu14} with a commitment time of 15ms. Note that this protocol involves quantum communications and measurements which greatly increase the complexity of security analysis theoretically \cite{Croke12}. Moreover, when implemented in the real world, one also need to consider the systematic errors in the preparation and transmission of the quantum particles by Bob, and also in the measurements executed by Alice. The purpose of this paper is to present a new relativistic bit commitment protocol which requires the exchange of classical information only. Due to the absence of quantum transmissions and measurements, this protocol can be shown to be unconditionally secure in a relatively transparent manner. Moreover it can also be readily implemented in realistic situations.

Let Alice controls two secure stations $(A_1, A_2)$, and Bob controls another two ($B_1, B_2$). $B_1$ and $B_2$ are separated by a large distance $d$, but the distance between $B_1$ and $A_1$ is negligible compared with $d$, so is that between $B_2$ and $A_2$. We assume that there are only classical communication channels between Alice and Bob, but the same cannot be guaranteed for internal communications between $A_1$ and $A_2$, or $B_1$ and $B_2$. We shall make the idealization that information transmits along all channels at the speed of light $c$. Since $A_i$ is adjacent to $B_i$ ($i=1,2$), for simplicity we shall take the time required to transmit information between them to be negligible, so is the time required to process information at each station. Let ($n_0, n_1, m_0, m_1, \eta$) be random but distinct bit-chains of length $l$ (i.e., they are numbers between 0 and $2^l-1$ in the binary form). It is understood that ($n_0, n_1$) are generated on demand by $B_1$, and ($m_0, m_1$) by $B_2$ independently. $\eta$ is a secret key shared by $A_1$ and $A_2$, it can be pre-established by any secure quantum key distribution protocol, such as BB84 \cite{BB84}. The new protocol can be specified as follows.
\begin{itemize}
\item[]\hspace{-0.5cm}{$Commit$:}
\item[1.]{Alice ($A_1$, $A_2$) decides the value of $b\in\{0,1\}$.}
\item[2.]{At time $t=0$, $B_1$ sends to $A_1$ two distinct random bit-chains $(n_0,n_1)$, and $A_1$ commits by returning the number $N_b=(\eta~ \textsc{xor}~n_b)$ to $B_1$. Independently also at $t=0$, $B_2$ sends $A_2$ two distinct random bit-chains $(m_0,m_1)$, and $A_2$ commits by returning $M_b=(\eta~\textsc{xor}~m_b)$ to $B_2$.}
\item[3.]{$B_1$ sends $(\lambda_0,\lambda_1)$ to $B_2$, where $\lambda_i=(N_b~\textsc{xor}~n_i)$. Similarly, $B_2$ sends $(\zeta_0,\zeta_1)$ to $B_1$, where $\zeta_i=(M_b~\textsc{xor}~m_i)$.}
\item[]\hspace{-0.5cm}{$Unveil$:}
\item[1.]{The protocol unveils when $B_1$ receives ($\lambda_0,\lambda_1$) and $B_2$ receives ($\zeta_0,\zeta_1$) at $t=d/c$ \cite{footnote}.}
\item[2.]{The committed value of $b$ is obtained from the conditions $\lambda_b=\zeta_b$ and $\lambda_{1-b}\ne\zeta_{1-b}$. If neither $b=0$ nor $b=1$ satisfies these conditions, then Alice is cheating.}
\end{itemize}

The unveiling mechanism can be understood as follows. Without loss of generality, suppose $A_1$ and $A_2$ commit to $b=0$, then $N_0=\eta~\textsc{xor}~n_0$ and $M_0=\eta~\textsc{xor}~m_0$.
It follows that
\beqa
\lambda_0=(\eta~\textsc{xor}~n_0)~\textsc{xor}~n_0=\eta,\\
\zeta_0=(\eta~\textsc{xor}~m_0)~\textsc{xor}~m_0=\eta,
\eeqa
so that $\lambda_0=\zeta_0$. However
\beqa
\lambda_1=(\eta~\textsc{xor}~n_0)~\textsc{xor}~n_1,\\
\zeta_1=(\eta~\textsc{xor}~m_0)~\textsc{xor}~m_1,
\eeqa
are uncorrelated random numbers, so that $\lambda_1\ne\zeta_1$. Nevertheless in principle one cannot strictly rule out the possibility that $\lambda_1=\zeta_1$ by chance, in which case the unveiled result is ambiguous. However the probability for this to happen is $1/2^l$, which can be made as small as desired by adopting a large $l$.

Next we show that this protocol is unconditionally secure. Clearly, $N_b$ and $M_b$ are respectively encryptions of $n_b$ and $m_b$, with $\eta$ being the common encryption key or one-time pad. It is well known in cryptography that, without the knowledge of $\eta$ and before $B_1$ and $B_2$ have time to exchange information, $N_b$ and $M_b$ are nothing but random numbers from which no information of $b$ can be extracted. Hence the commitment is concealed for $0\le t< d/c$. Here we have implicitly assume that Bob sets up only two stations ($B_1$ and $B_2$) as stipulated by the protocol. However in principle nothing can prevent him from setting up a secret station $B_3$ somewhere else. The optimal location for $B_3$ is half way between $B_1$ and $B_2$. Then the information sent by $B_1$ and $B_2$ at $t=0$ (step 3 of $Commit$) will both arrive $B_3$ at $t=d/2c$, and the commitment is unveiled. Therefore, for unconditional security, the length of the commitment period ($\Delta t$) should be taken to be $d/2c$ instead of $d/c$.

Next we consider the issue of binding. As we saw, $A_1$'s commitment is separated from $A_2$'s by a space-like interval, the principle of relativity dictates that these two events are causally independent. Hence a dishonest Alice has only two choices before making the commitment: (1) Do $A_1$ and $A_2$ commit to the same $b$? and (2) Do they use the same $\eta$? Now since $A_1$ and $A_2$ must commit at the same moment $t=0$, it does not make sense logically for them to commit to different values of $b$. It is easy to show that, if $A_1$ and $A_2$ commit to different values of $b$, then the conditions $\lambda_b=\zeta_b$ and $\lambda_{1-b}\ne\zeta_{1-b}$ will not be satisfied by neither $b=0$ nor $b=1$, independent of whether they use the same $\eta$ or not. Given that $A_1$ and $A_2$ commit to the same $b$, they must use the same $\eta$, otherwise the unveiling process will again be unsuccessful, even if they choose to be honest. The argument is more transparent and straightforward from the following point of view. Notice that the protocol is completely out of Alice's control after she commits and she does not play any role in the unveiling phase. This fact already implies that Alice cannot cheat. The reason is that, no matter which value of $b$ is revealed in the unveiling phase, it is a direct consequence of what Alice did when she committed, not what she might have done afterward. We conclude that the protocol is safe against classical attacks.

We now consider quantum attacks. As mentioned before, all the information exchanged between Alice and Bob in the commitment phase are restricted to be classical in nature. That is, the numbers $n_0,n_1,m_0,m_1,N_b,$ and $M_b$ are all concrete classical numbers without any ambiguities to the respective receivers. Bob has no quantum cheating strategy because he must send classical numbers ($n_0,n_1,m_0,m_1$) to Alice, and also receives classical numbers ($N_b,M_b$) in return. Alice does not have much freedom either because the numbers $N_b$ and $M_b$ she sends to Bob in the commitment phase are restricted to be classical. It is well known that quantum attacks depend on the fact that the dishonest party can leave all undisclosed parameters undetermined at the quantum level \cite{LoChau97,Mayers97}. Now the only parameter that Alice keeps secret is $\eta$. Although $\eta$ is not known to Bob before unveiling, it is clear that Alice cannot leave it undetermined at the quantum level when she commits, otherwise she won't be able to produce the two classical numbers $N_b$ and $M_b$ to be sent to Bob. This point is also obvious from the fact that Alice does not play any role in the unveiling phase of protocol, that is, she is not required to specify the unknown parameter $\eta$ when the protocol unveils. This fact clearly implies that she has no freedom to leave it undetermined when she commits. We conclude that neither Alice nor Bob has a viable quantum cheating strategy.

Furthermore since Bob sends to Alice classical descriptions of ($n_0,n_1$) and ($m_0,m_1$), and she must send back classical numbers ($N_b,M_b$) in return, Alice can only make classical commitments. That is, she has no freedom to commit to a superposition of $b=0$ and $b=1$. That means when our protocol is used as a subprotocol in a more complicated task, one can be sure that $b$ has a definite classical value before the protocol is opened \cite{Kent12}.

For simplicity, we have made the idealization that information are transmitted at light speed $c$, and that the distance between $A_i$ and $B_i$ is negligible, so is the time required for the agents to process information. There is no difficulties to take into account the actual limitations when implemented in the real world. Realistically we can specify that Alice's commitments must be completed at $t=\delta<< d/c$. The length of the commitment time remains $\Delta t=d/2c$, starting from $t=\delta$ until $t=\delta+d/2c$. The actual speed of information transmission between $B_1$ and $B_2$ may be less than the speed of light, but Alice cannot be sure of that, so $\Delta t=d/2c$ can be regarded as an unconditionally safe lower bound.

Notice that in the protocol proposed above, $A_1$ and $A_2$ play symmetric roles, so do $B_1$ and $B_2$. However in some applications it may be more convenient or advantageous to have one of Alice's and Bob's stations (say $A_2$ and $B_2$) play subordinate roles only. Such a protocol is also possible and can be specified as follows.

\noindent $Commit$: (1)$A_1$ decides the value of $b\in\{0,1\}$.
(2)At time $t=0$, $B_1$ sends to $A_1$ two distinct random bit-chains $(n_0,n_1)$, and $A_1$ commits by returning the number $N_b=(\eta~\textsc{xor}~n_b)$ to $B_1$. At the same time ($t=0$), $A_2$ sends $\eta$ to $B_2$, and $B_2$ sends $\eta$ to $B_1$.

\noindent $Unveil$: (1)The protocol unveils when $B_1$ receives $\eta$ at $t=d/c$ from $B_2$ \cite{footnote}. (2)The committed value of $b$ is obtained from the condition $n_b=\eta~\textsc{xor}~N_b$.

In this protocol, $A_2$ does not need to know the value of $b$ to be committed, her only role is to deliver the encryption key $\eta$ to $B_2$. From the discussions we have gone through earlier, the security of this alternative protocol is quite obvious: (1) $B_1$ cannot extract any information from $N_b$ until he receives the key $\eta$ from $B_2$. (2) Alice cannot cheat because the protocol is out of her control after she commits. (3) The unconditionally secure commitment time is again $\Delta=d/2c$ as before.

To summarize, we have proposed a new unconditionally secure bit commitment protocol base on the relativistic restriction that signals cannot travel faster than the speed of light. This protocol does not involve any quantum particles so that the theoretical security analysis is relatively simple and transparent, moreover it can be readily implemented experimentally. Alice commits by making two encryptions separated by a space-like interval. Security against Bob is guaranteed by the fact that an encrypted message contains no extractable information without the key. One special feature of the protocol is that Alice does not play a role in the unveiling phase, in fact she has no control over the protocol after she commits; this feature clearly implies that Alice cannot cheat by classical or quantum means. In addition to the original one, we have also proposed an alternative protocol where one of Alice's stations plays only a subordinate role. This protocol is also unconditionally secure, and it may be more convenient or advantageous in certain cryptographic applications.





\begin{thebibliography}{}
\bibitem{footnote} For unconditional security, the protocol should be taken to be unveiled at $t=d/2c$. See discussions in the text.
\end{thebibliography}
\end{document}